# Point-contact enhanced superconductivity in trigonal $PtBi_2$: quest for the origin of "high-$T_c$"

O. E. Kvitnitskaya[1,2], L. Harnagea[2,3], G. Shipunov[2], S. Aswartham[2*], I. Kovalchuk[4],
V. V. Fisun[1], D. V. Efremov[2], B. Büchner[2,5], Yu. G. Naidyuk[1]

[1]*B. Verkin Institute for Low Temperature Physics and Engineering, NAS of Ukraine, 61103 Kharkiv, Ukraine*

[2]*Leibniz Institute for Solid State and Materials Research, IFW Dresden, 01069 Dresden, Germany*

[3]*I-HUB Quantum Technology Foundation, Indian Institute of Science Education and Research (IISER), Pune 411008, India*

[4]*Kyiv Academic University, 03142 Kyiv, Ukraine*

[5]*Institute of Solid State and Materials Physics and Würzburg-Dresden Cluster of Excellence ct.qmat, Technische Universität Dresden, 01062 Dresden, Germany*

## Abstract

We studied enhanced superconductivity in point contacts (PCs) based on a type-I Weyl semimetal trigonal *t*-$PtBi_2$ using both normal metal (Ag, Cu, Pt) and ferromagnetic (Fe, Co, Ni) tips by measuring the differential resistance *dV/dI(V)* curves. In most cases, the value of the superconducting critical temperature $T_c$ ranges between 3 and 5 K, which is several times higher than the maximum bulk $T_c$. Notably, among the various PCs we examined, a few achieved $T_c$ values as high as 8 K, including those with both normal and ferromagnetic tips. Additionally, the critical magnetic field is also highly enhanced in these PCs and reaches up to several Tesla. The common reason for the $T_c$ increase may be related to pressure/strain caused during the PC's formation. It is worth noting that a greater increase in $T_c$ is observed in PCs formed at the edge of the sample flake, compared to those formed on the plane of the platelet. The results also reveal that the enhancement of $T_c$ in PCs based on *t*-$PtBi_2$ is compatible with ferromagnetic tip, which may suggest a potentially complex nature of enhanced superconductivity. Our findings besides suggest that *t*-$PtBi_2$ is a promising candidate for realizing topological superconductivity at more accessible temperatures.

***Present address:** International Research Centre MagTop, Institute of Physics, Polish Academy of Sciences, al. Lotników 32/46, 02-668 Warsaw, Poland.

## Introduction

Two-dimensional materials are at the forefront of fundamental and applied research because of their wide range of electronic ground states. When scaled to monolayers and atomic dimensions, they exhibit new quantum phenomena. One example is trigonal t-$PtBi_2$, recently classified as a type-1 Weyl semimetal with a superconducting transition below 1K [1], making t-$PtBi_2$ an attractive candidate for realizing topological superconductivity and to create sustained qubits. t-$PtBi_2$ gained researchers' attention due to the observation of robust low-dimensional superconductivity [2], indicating superconductivity in a type-I Weyl semimetal. Additional motivation for exploring superconductivity in t-$PtBi_2$ and its surface peculiarities came from point-contact (PC) measurements [3]. These latter show an increase in the superconducting critical temperature $T_c$ in PCs, reaching up to 3.5 K, which is about ten times higher than the bulk $T_c$. Other remarkable findings include tunneling measurements, which detected extremely large gap values and a quasiparticle excitation spectrum resembling that of high-$T_c$ superconductors [4]. ARPES studies further revealed topological Fermi arcs on opposite surfaces of non-centrosymmetric t-$PtBi_2$, which are still superconducting around 10 K [5, 6]. The agreement between experimental quasiparticle interference patterns and first-principles calculations of the topological Fermi arc joint DOS [7] supports the dominance of these arcs in the surface electronic properties of t-$PtBi_2$, including the observed surface superconductivity [4-6]. Nonetheless, some aspects of t-$PtBi_2$'s unusual superconducting behavior remain unclear. These include the nonhomogeneity of the superconducting state [8], the absence of superconducting signatures in low-temperature specific heat and magnetization data [4, 8], the lack of vortices [4], the failure to observe superconductivity down to 3 K in topography of Fermi arcs according to ARPES measurements [9], controversial observation of vortex lattice and robust surface superconductivity with $T_c$ = 2.9 K [10], and visualization of both bulk ($T_c$ ~0.5K) and at certain surface regions ($T_c$ ~3 K) superconductivity using ultra-low-temperature scanning tunneling spectroscopy at 5 mK [11]. Consequently, we present here further comprehensive PC measurements based on extensive statistical data to help our understanding of the origin of "high-$T_c$" superconductivity in t-$PtBi_2$.

## Experimental details

We used samples that were grown in different batches in PC experiments. Detailed information on sample growth is given in the Supplement and Refs. [1, 12]. Some crystals were cleaved into thinner pieces and create the PCs on freshly exposed surfaces. In general, there was no significant difference between the data collected from fresh cleaved surfaces and that from the as-grown flux-free crystal surfaces. Additionally, no qualitative differences were found among the data gathered from different crystals, with various RRR values ranging from 130 to 400. This suggests that the quality of the crystals is not the primary issue. Instead, the more significant factor is the "quality" of a specific PC, which is hard to control due to its submicron size and "handmade" technique used in its preparation. As with many experimental techniques, the reproducibility of the collected data and its statistical processing should always take precedence.

"Hard" or "clamping" type PCs were made directly in a helium cryostat by mechanically touching the wire of normal (Cu, Ag, Au), or ferromagnetic metals (Fe, Co, Ni) to the edge or to the plane of a flake. "Soft" contacts were made by placing a small drop of a conductive Ag paint between the sample and a 50 μm copper wire. Thus, "soft" contacts were prepared at ambient atmosphere and temperature and moved into the cryostat after the silver paint had dried out. In this study, we also paid more attention to the measurement of homocontacts by touching two flakes of t-$PtBi_2$ edge to edge or edge to plane. Thus, we conducted resistive measurements on the hetero- homo- and "soft"- contacts based on t-$PtBi_2$ using normal and magnetic counter-electrodes.

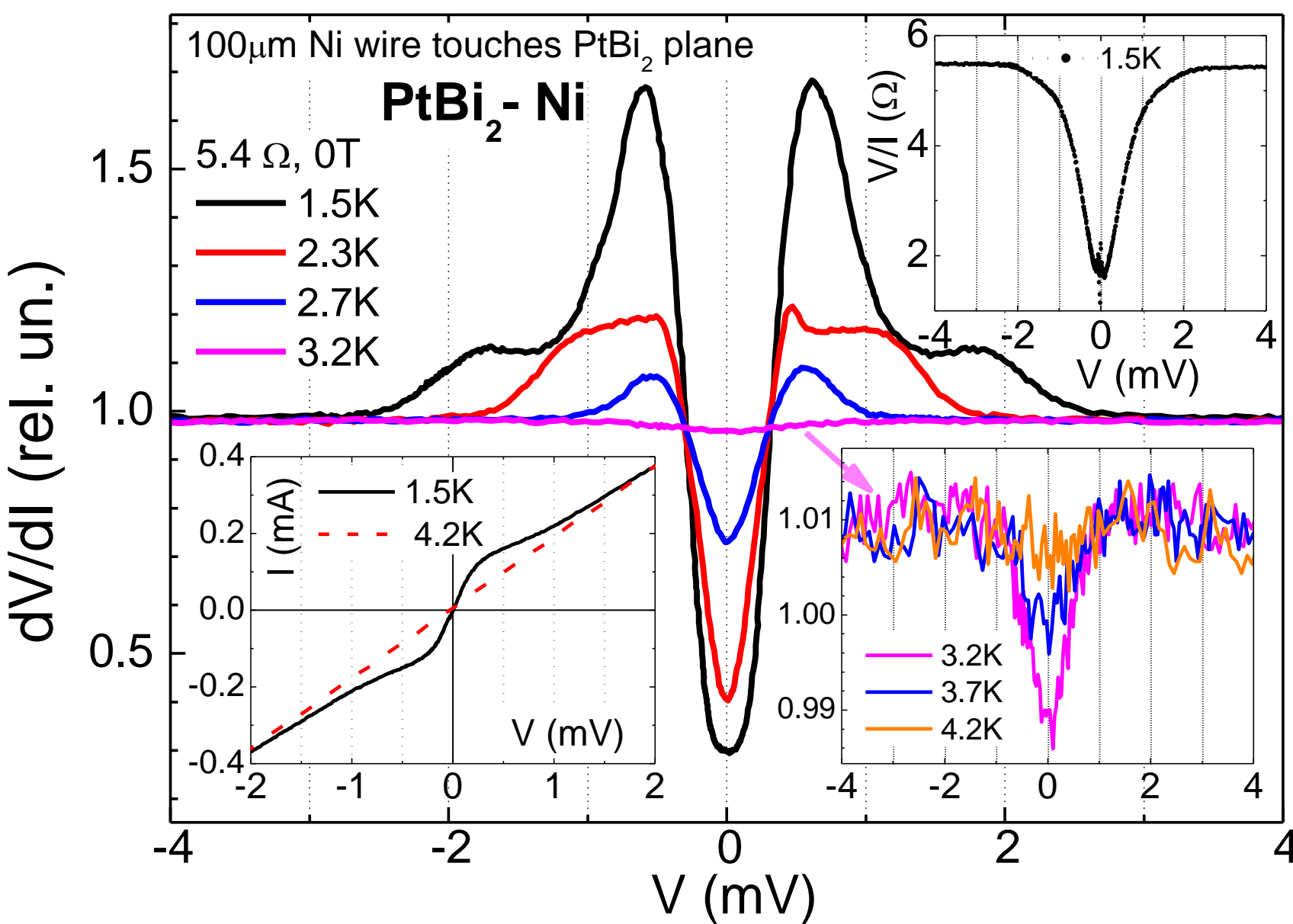

**Fig. 1.** Temperature variation of *dV/dI* of $PtBi_2$ – Ni PC with resistance R = 5.4 Ω. The left inset shows the current-voltage *I(V)* curve in normal (dashed red) and superconducting state (solid black). The upper inset shows the calculated static PC resistance *V/I* at 1.5 K. Bottom inset on the right shows the behavior of residual superconducting minimum as PC approaches the normal state.

We measured the current-voltage *I(V)* characteristics of PCs and their first *dV/dI* derivative by the standard look-in technique. The $dV/dI(V) \equiv R_d(V)$ was recorded by sweeping the *dc* current *I* on which a small *ac* current *i* was superimposed. The measurements were performed in the temperature range between 1.5 and 10 K and in a magnetic field up to several tesla.

## Results

Figure 1 shows the most typical *dV/dI* curves from more than one hundred and fifty studied PCs. The variety of other *dV/dI* spectra of $PtBi_2$ heterocontacts is shown in Fig. S2 in the Supplement. The shape of *dV/dI* in Fig., characterized by a deep zero-bias minimum and symmetrical side maxima, differs from the typical Andreev reflection features, which usually display symmetric versus V=0 double minima structure in the region of the superconducting gap [13]. Considering the variations of the *dV/dI* curves shown in Fig. S1 in the Supplement, we believe that the *dV/dI* patterns are because of critical current and/or heating effects due to an increase in voltage [13, 14]. Consequently, obtaining spectral information, such as details regarding the superconducting gap, becomes challenging. Therefore, the most reliable parameters we can extract from these measurements are the critical temperature ($T_c$) or the critical magnetic field ($B_c$), which we estimate at the point when all superconducting features in the dV/dI curve disappear. As a result, this paper will primarily focus on the behavior of $T_c$, based on extensive statistical data collected from various $PtBi_2$ samples and electrodes (tips) made from both normal and magnetic elemental metals.

Figure 2 shows the distribution of $T_c$ for all measured heterocontacts based on $PtBi_2$ with normal (Cu, Ag, Pt) and ferromagnetic metals (Fe, Co, Ni). The enhanced $T_c$ primarily falls between 3 and 5 K, which aligns with our previous studies [3]. However, numerous PCs exhibit $T_c$ values exceeding 5 K, with some reaching as high as 8 K (see Fig. S1(d) in the Supplement). This is more than double the previously reported values [3, 15], and has been observed in various PCs using both Cu and magnetic Ni tips. Thus, we conclude that the enhanced superconductivity in t-$PtBi_2$ PCs is unaffected by the presence of ferromagnetism in the counter-electrode (tip). Additionally, the critical magnetic field has also been significantly increased, reaching up to 3 T, which is consistent with past reports [3, 15].

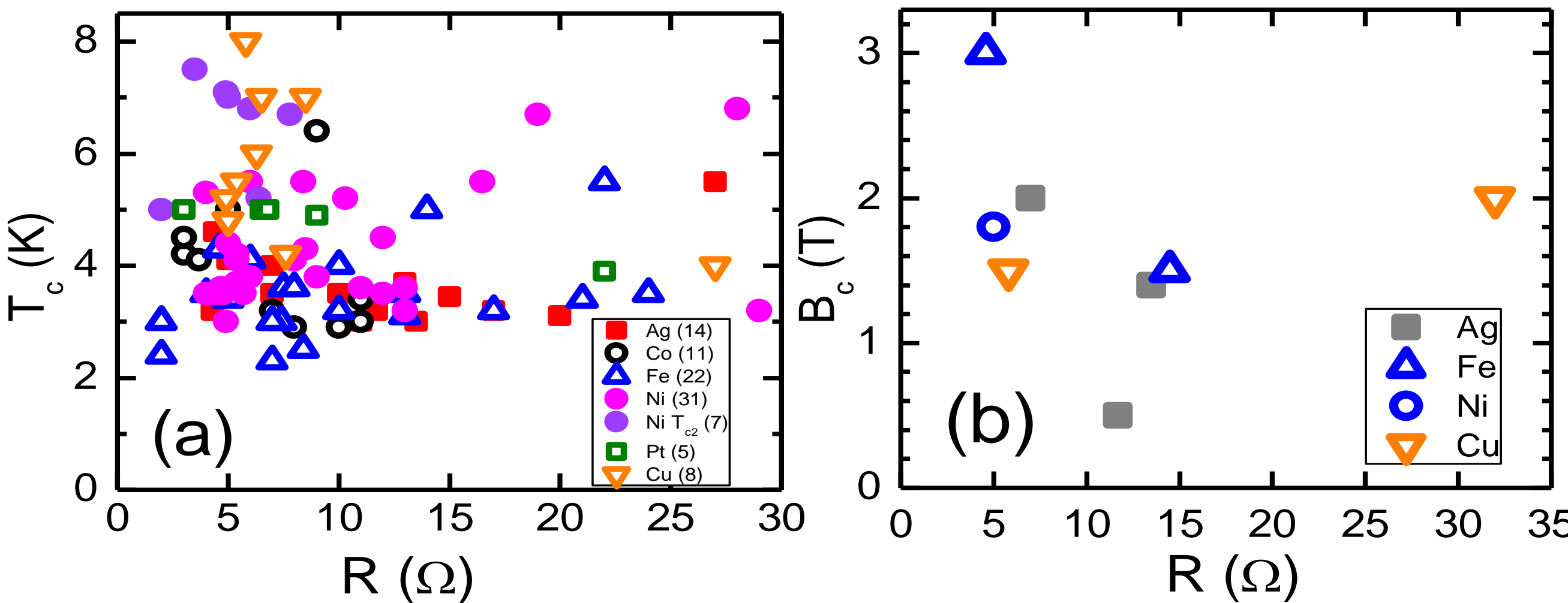

**Fig. 2**. (a) $T_c$ distribution for all measured heterocontacts with normal (Ag, Cu, Pt) and ferromagnetic (Fe, Ni, Co) metals (tips). (b) Critical magnetic field for some PCs with normal and ferromagnetic electrodes at temperatures between 1.6 and 1.9 K. The x-axis shows PC resistance in the normal state.

Figure 3 (a, b) demonstrates histograms of $T_c$ distribution for all measured heterocontacts with normal and magnetic electrodes, separated on PCs made by touching with a metal wire the plane of $PtBi_2$ flake (a) or the edge (b). It is seen that in average $T_c$ for “edge” PCs is 0.9 K larger compared to the “plane” PCs. Figure 3 (c, d) demonstrates the same for PCs with ferromagnetic Ni, for which we have more extensive statistical data. Here, on average, $T_c$ is almost the same as for all PCs from the left panel, and the same correlation is observed here, specifically $T_c$ for “edge” PCs is larger than that for the “plane” PCs.

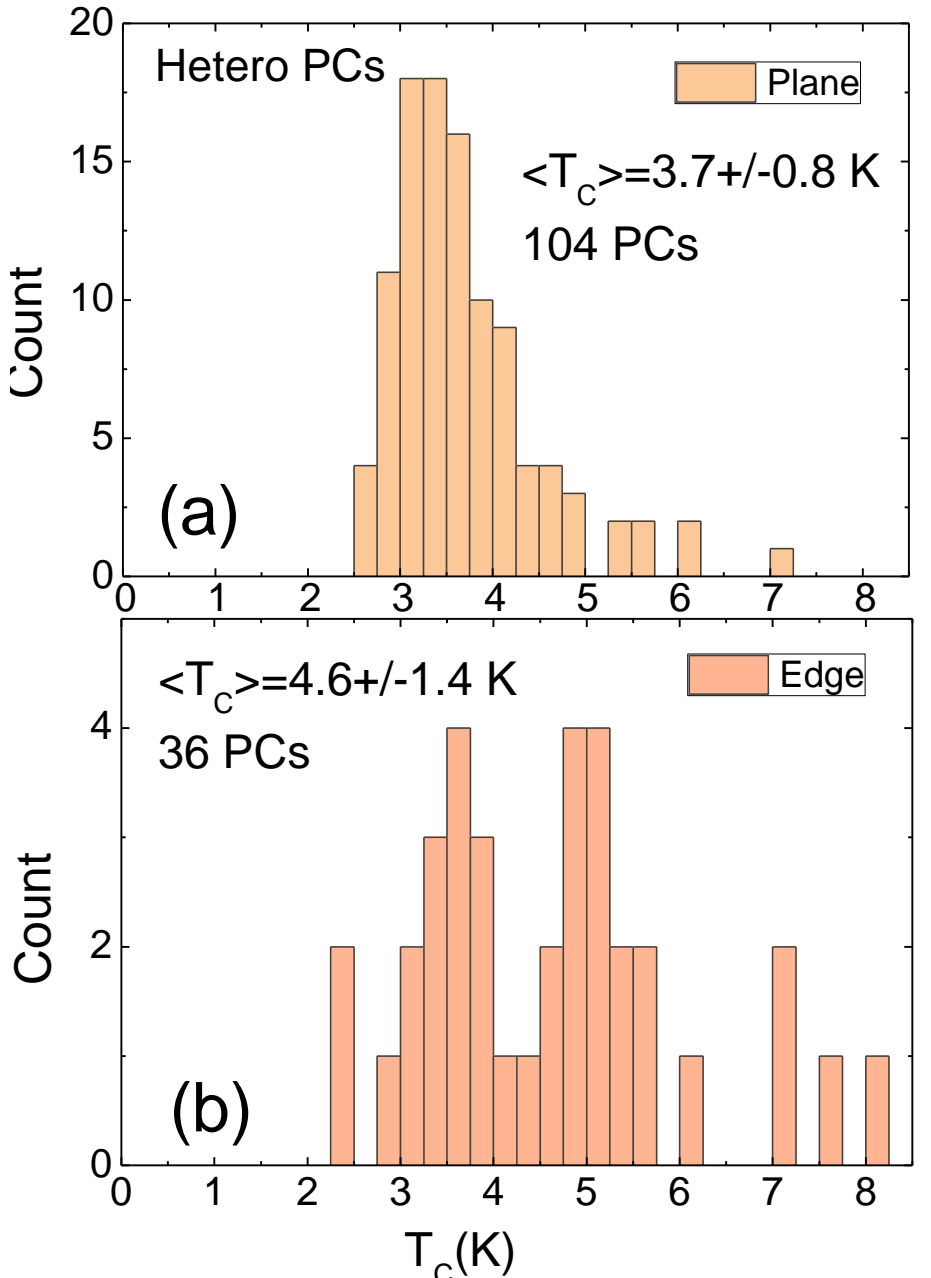

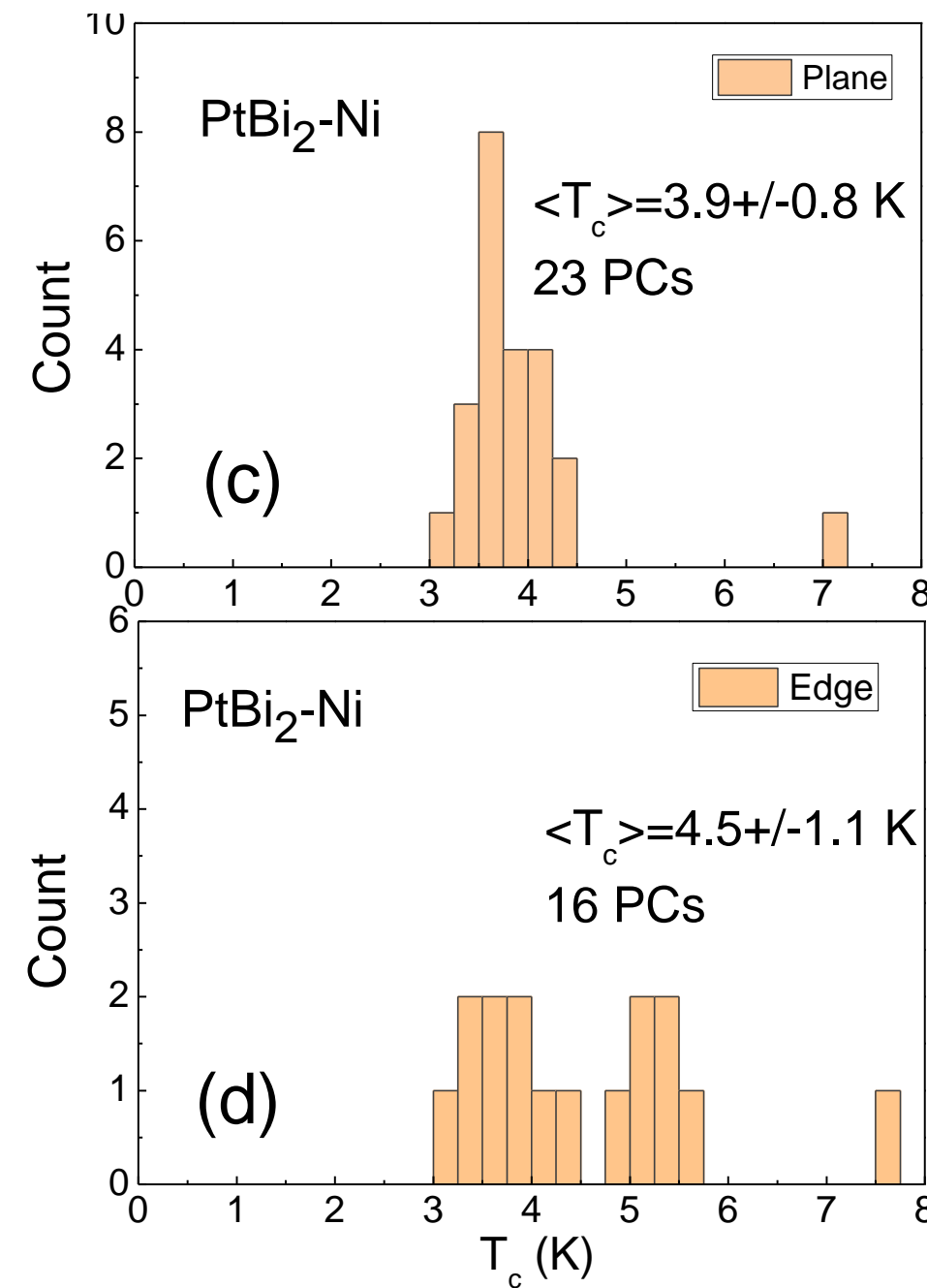

**Fig. 3**. (a, b) Histogram of $T_c$ distribution for all measured $PtBi_2$ heterocontacts with normal (Ag, Cu, Pt) and ferromagnetic (Fe, Ni, Co) electrodes made by the touching of a metal wire to the plane of $PtBi_2$ (a) or to the edge (b). (c, d) Similar histogram for $PtBi_2$ heterocontacts with ferromagnetic Ni.

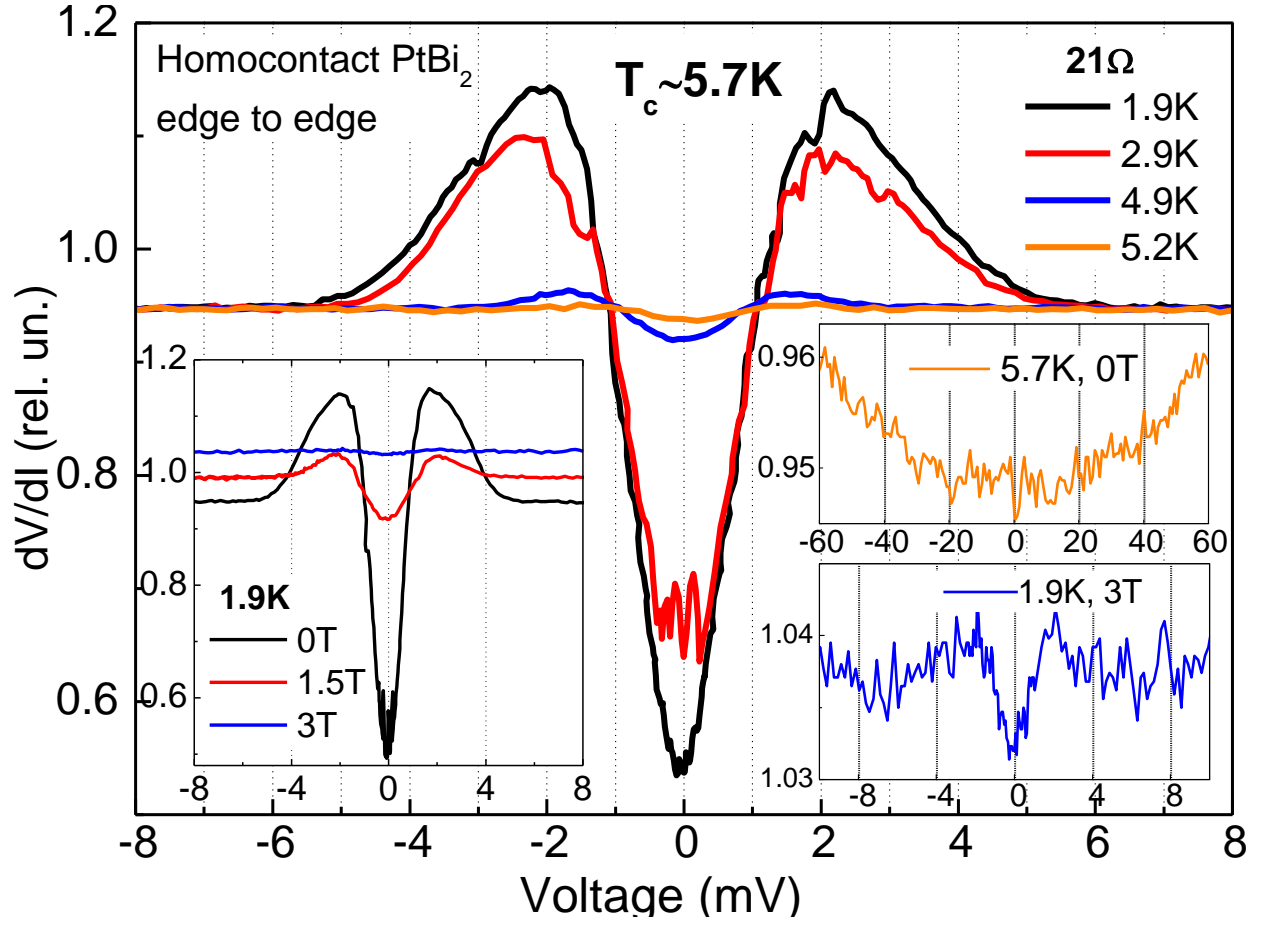

**Fig. 4.** Temperature variation of *dV/dI* of $PtBi_2$ – $PtBi_2$ PC with R = 21 Ω. The left inset shows *dV/dI* in the magnetic field at T = 1.9K. The right inset shows *dV/dI* behavior close to the $T_c$ (upper panel) or near the critical magnetic field (bottom panel).

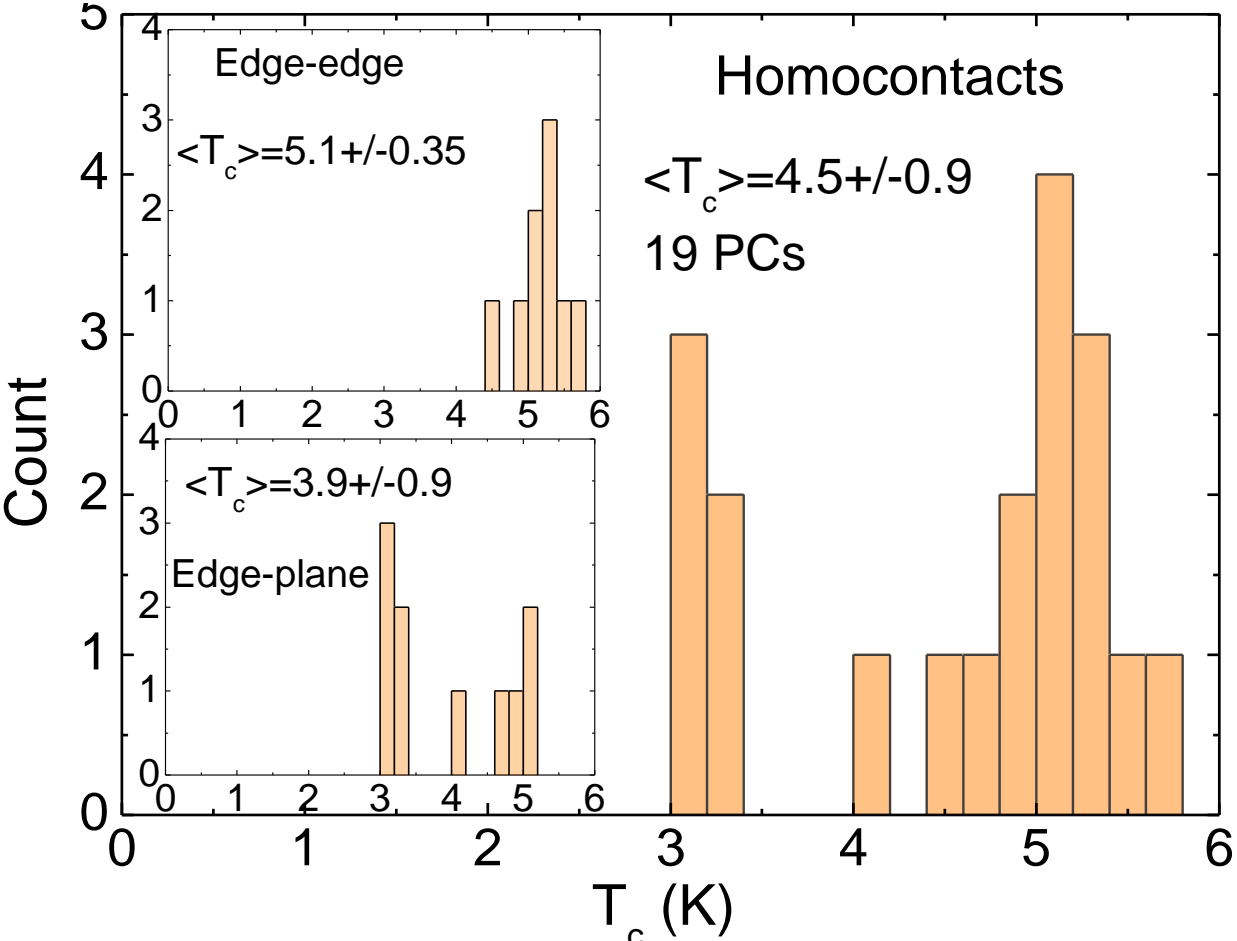

**Fig. 5**. Histogram of $T_c$ distribution for all measured $PtBi_2$ homocontacts. Insets show histograms separately for homocontacts made by touching edge to edge and edge to plane.

Figure 4 demonstrates the typical temperature and magnetic field behavior of *dV/dI* for $PtBi_2$ homocontacts. In general, its shape is similar to the case of heterocontacts. Note that there is no supercurrent (critical current) at V=0, and PC has a nonzero resistance. This means that the PC is only partially in a superconducting state. So far, it looks typical for enhanced superconductivity, e.g., the same nonzero resistance we observed in the $MoTe_2$ homocontacts [16].

Figure 5 shows histograms of $T_c$ distribution for all measured homocontacts, which also separated on PCs made by contacting the plane of $PtBi_2$ (a) or the edge (b). It is seen that in average $T_c$ for "edge" PCs is 1.2 K larger compared to the "plane" PCs. Also, the averaged $T_c$ for homocontacts is greater than the same in heterocontacts.

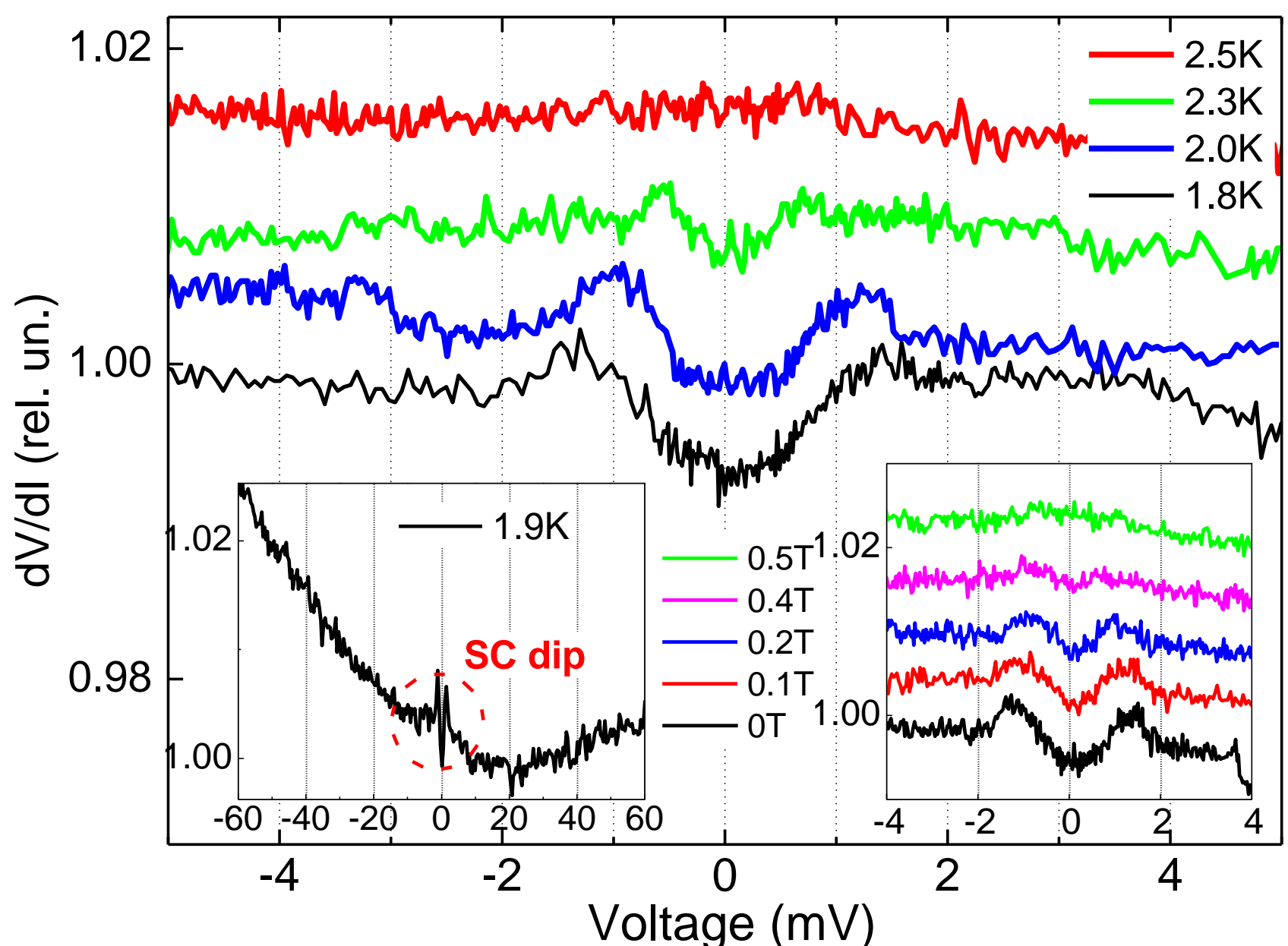

**Fig. 6.** Temperature variation of *dV/dI* of $PtBi_2$ "soft" PC with resistance R = 1.4 Ω in the normal state. Left inset shows *dV/dI* in the extended voltage range at T = 1.9K. The right inset shows the *dV/dI* behavior in a magnetic field at T = 1.9 K.

Figure 6 displays the temperature and magnetic field behavior of *dV/dI* for $PtBi_2$ "soft" contacts. All "soft" contacts were prepared on the edge, because contacts to the plane were unstable and fell off. In general, only about ten of the "soft" contacts were studied, and only about half of them exhibited weak features of superconductivity, with the depth of the zero-bias minima being approximately a few percent.  Nevertheless, the $T_c$ was not radically low, since its variation for "soft" contact was between 2.5 and 4.5 K, which corresponds to the left wing of the histograms for "hard" PCs, and the critical magnetic field was in the range 0.5 - 2 T.

## Discussion

The study of superconductivity using PCs began to advance rapidly in the early 1980s through the exploitation of phenomena such as Andreev reflection (see Ref. [13] and references therein). Initially, the first publications appeared [17, 18], where an increase in $T_c$ was observed in PCs with elemental metals such as Ga and Bi. Since then such phenomena have been observed and investigated for more than twenty different elements and compounds. Among them are doped semiconductors Si and Ga [19, 20], heavy fermion compound $CeCu_2Ge_2$ [21], iron-based superconductors FeSe and $A Fe_2As_2$ ($A$ = K, Cs, Rb) [22, 23], different topological and Weyl semimetals [24-30], candidate to spin-triplet pairing $Sr_2RuO_4$ [31], Kagome Compounds such as $KV_3Sb_5$ and $CsV_3Sb_5$ [32] along with several others. The enhancement of $T_c$ was primarily attributed to localized high pressure or mechanical strain exerted by a sharpened tip on the sample surface. This modification alters the crystal and electronic structure, boosting the superconducting properties. The literature discusses several potential origins for the enhanced $T_c$, including those induced by structural phase transitions at the tip, unique characteristics of the surface or interface, and the disorder or amorphous state created in the contact region. Below, we will explore these possible origins in relation to our measurements with t-$PtBi_2$.

*Origin of enhanced $T_c$.* It is well known that the pressure, as a rule, increases the critical temperature in superconductors (see, e.g., [33]). The role of pressure and deformation was considered already in the first experiments, where enhanced $T_c$ in PCs based on elemental metals Ga [17] and Bi [18] was observed. So-called pressure-type or "hard" PCs, created by a tip touching a sample, can produce significant pressure due to the small contact area. The pressure in PC is challenging to estimate; however, it is fundamentally limited by the stiffness of the materials in contact. The pressure origin of enhanced $T_c$ in PCs is supported by the fact that usually $T_c$ rises under pressure in the same bulk material as well. So, pressure plays a role in our case, however, we remind that the $T_c$ in bulk $PtBi_2$ increases only up to 2 K under pressure of 5 GPa and keeps this value up to 48 GPa [34]. That is, we observed significantly larger $T_c$ in PCs, than what ordinary hydrostatic pressure can produce. Probably, the nonuniform distribution of pressure in clamped "hard" PC plays a role in our case, especially considering the observed difference in $T_c$ between "edge" and "plane" created PCs. In any case, this difference suggests the need to consider a more sophisticated origin of enhanced $T_c$.

*Pressure/strain in "soft" PCs.* The size of the "soft" PC, made through a drop of silver paint, is approximately tenths of a millimeter, and the PC consists apparently of tens or more parallel nanoscale conducting channels between individual silver particles and the sample surface, as demonstrated by a scanning electron microscope in Ref. [35]. No pressure/strain is expected unlike in the case of mechanically prepared PCs. However, some disturbance of the surface layers can occur, e.g., after solidification of the silver paste or upon cooling due to the different thermal expansion between the sample and the silver paste, as noted in Ref. [32]. In such cases, only the topmost layers will be affected, which produces a minor contribution to the PC resistance. This is in line with the shallow depth of the zero-bias minimum of the order of a few percent in "soft" PCs (see Fig. 6) compared to "hard" PCs. On the other hand, "soft" PCs are more sensitive to surface properties, and they may reflect more surface superconductivity observed by STS [4, 10, 11] and ARPES [5,6], which is likely non-homogeneous and of an insular type.

*Multiple $T_c$.* Another interesting observation is that the shape of some *dV/dI* curves resembles a superposition of two superconducting minima (see Fig. S1e). If such PCs were in the spectroscopic regime with characteristic Andreev reflection features, i.e., structures with double minima, as on the *dV/dI* of typical two-gap superconductor $MgB_2$ [36], we could speculate about two-gap superconductivity. However, it seems that our PCs are in a nonspectroscopic regime. Similar characteristics have frequently been observed, for example, in PCs based on FeSe [22] and other iron-based superconductors [37]. In these cases, all superconducting peculiarities in *dV/dI(V)* vanished at the same temperature as it increased. In the case of *t*-$PtBi_2$, the narrow zero-bias minimum disappears first at a temperature of about 3 K (see inset in Fig. S1e), while the broad one vanishes at higher temperatures close to 7 K. Probably, such characteristics are related to the distribution of $T_c$ and can be obtained in the case of multicontact formation.

*Absence of typical Andreev reflection features*. Despite conducting numerous measurements of *dV/dI* for over one hundred and fifty PCs, we never observed curves with the typical double-minimum structure characteristic to Andreev reflection. This structure was observed in $MoTe_2$ [16] and in $PtBi_2$ by Di et al. [15], although the latter was only measured with a nonmagnetic electrode. The reason may be that our measurements were restricted to a temperature of 1.5 K, while the lowest temperature was 0.7 K in Ref. [15]. Another crucial reason is the short coherence length $\xi$ of approximately 5-10 nm, as evidenced by STS data [4, 10], which can be significantly smaller than the PC size d. At the same time, $\xi$ must be larger than the contact size d due to the requirements of the theory by considering Andreev reflection in PCs. In our case, the estimation of the PC size yields values between 45 and 145 nm for PC resistance between 3 and 30 Ohm (see Supplement). If $d \gg \xi$, the order parameter or gap may be distributed or variable within the PC core, resulting in an unpredictable *dV/dI* shape.

*PCs with magnetic tips.* We did not observed essential difference in $T_c$ increase between normal and magnetic tips, as indicated by our statistical data for $T_c$ in Figs. 2 and 3. The ferromagnetic state in the tip does not appear to have any effect on enhanced superconductive properties. This is consistent with the results of Ref. [38], where a detailed study of suppression of superconductivity in PCs between iron and elemental superconductor was carried out. The authors conclude that PCs with the strong superconductors Nb and Ta appear to be barely affected by using of Fe as a tip, whereas PCs with weaker superconductors, such as Sn and Al with a large coherence length, have a reduced $T_c$. The authors proposed that this is related to the magnetic field of about 10 mT generated by the ferromagnetic electrode. Taking into account that the critical field in our PCs is above a few Tesla, it is hard to expect the influence of the magnetic tip on the superconducting state in our PCs from the point of view of Ref. [38]. Anyway, it needs to be clarified whether the "unconventional" surface superconductivity in t-$PtBi_2$, which may have exotic i-wave origins [6], will compete or coexist at the ferromagnetic interface.

## Conclusions

1. Pressure/strain arising at the mechanical creation of “hard” PCs is the primary origin of enhanced $T_c$. At the same time, the enhanced $T_c$ is more than two times larger in comparison to measurements under hydrostatic pressure [34]. Unidirectional or non-uniformly distributed pressure/strain may play a role here.

2. Enhanced $T_c$ is, on average, greater for PCs prepared on the edge of flakes compared to the plane-contacted PCs. This naturally implies the use of non-hydrostatic pressure, bearing in mind that deformation along the plane can lead to an greater increase in $T_c$.

3. There is no essential difference in $T_c$ enhancement by using normal or ferromagnetic metals as counter electrode (tip). Nevertheless, this compatibility of enhanced superconductivity with magnetism may suggest a potentially complex nature of enhanced superconductivity.

4. Observation of statistically the same enhancement of $T_c$ both in homo- and heterocontacts eliminates "doping" as a possible reason of $T_c$ rise in the case of heterocontacts.

5. "Soft" PCs typically exhibit weaker signs of superconductivity with a lower $T_c$ ranging from 2.5 to 4.5 K. These observations align with the predominantly pressure/strain impact on enhanced $T_c$ in "hard" PCs. Conversely, the delicate disturbance produced by "soft" PCs or its absence may serve as confirmation of the enhanced "surface" superconductivity detected by STS [4, 10, 11] and ARPES [5, 6] techniques that probe only the topmost layers of the materials.

**Acknowledgement**

We are grateful to S. Gaβ and T. Schreiner for the technical assistance. We appreciate the insightful discussions with Mahdi Behnami about the transport and magneto-transport measurements on t-$PtBi_2$. We would like to acknowledge funding from the Alexander von Humboldt Foundation. OK, DE and BB acknowledge DFG Grant (BU 887/31-1). OK and YuN are also grateful for support by the NAS of Ukraine under project Ф19-5 and thankful to the IFW Dresden for hospitality. LH acknowledges support from IFW Dresden and I-HUB QTF, Pune. The work of IK was supported by the German Federal Ministry of Research, Technology and Space (BMFTR) through the GU-QuMat project (01DK24008).

**SUPPLEMENT**

to the paper by O. E. Kvitnitskaya et al. " Point-contact enhanced superconductivity in trigonal $PtBi_2$ ......"

**1. Sample grows and characterization.**

Single crystals of t-$PtBi_2$ were grown using the self – flux method, starting from various molar ratios of ultra-pure elemental Pt (4N, Evochem, Advanced Materials) and Bi (5N, chemPur), utilized as pieces or in powder form as explained in detail in Refs. [1, 12]. The elements were typically placed into a quartz ampoule, evacuated and sealed under high vacuum, melted and homogenized at 800 ºC, and the crystals were grown above 420 ºC at cooling rates as low as 1 to 2ºC per hour. The ampoule was promptly quenched in cold water after the flux was extracted using a centrifuge at this temperature. X-ray diffraction, scanning electron microscope (SEM) equipped with energy dispersive X-ray (EDX) probe, transmission electron microscope (TEM), magnetic and magnetoelectric measurements, etc. were used to thoroughly characterize the as – grown single crystals (see Fig. 1S). The obtained single crystals vary in size from a few to 10 mm, and they are pure, stoichiometric, well-crystallized platelets that can be easily cleaved along the ab-plane. Overall, the results align with earlier publications.

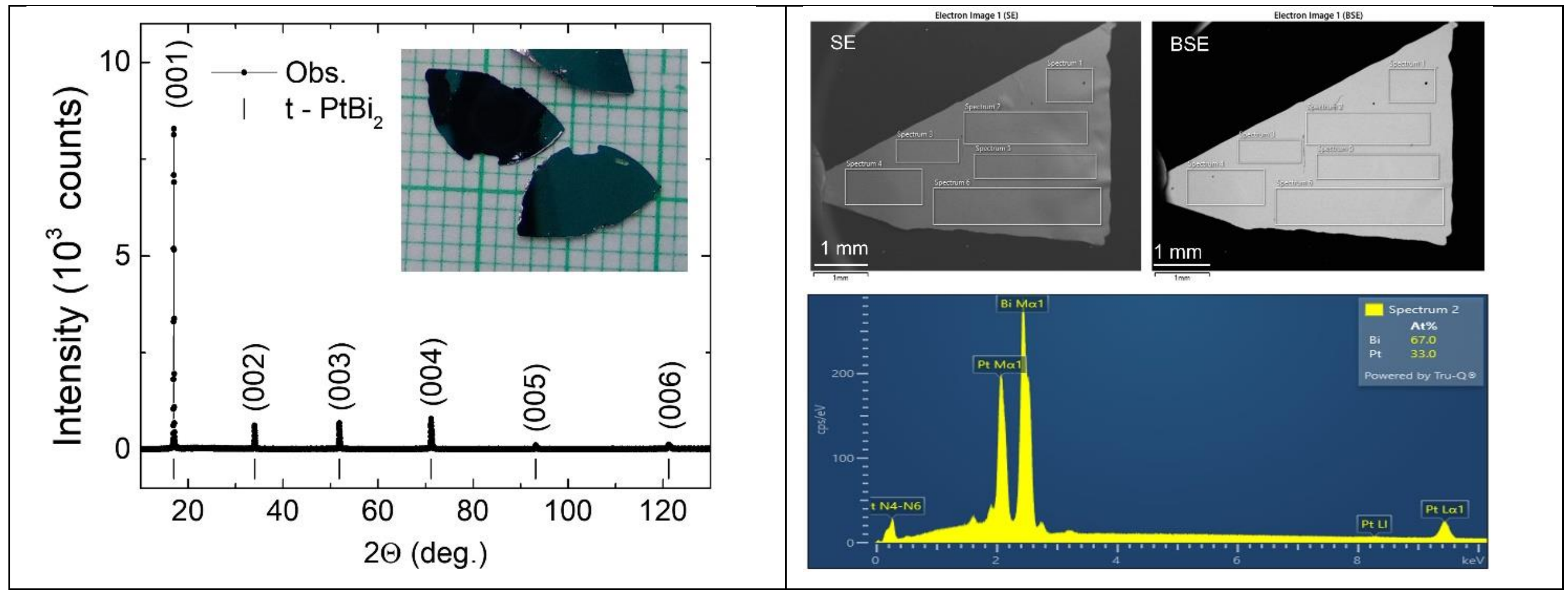

**Fig. S1**. Left panel: X-ray diffraction pattern measured in a Bragg-Brentano geometry (using Co K-alpha radiation) along the [001] direction on a single crystalline platelet of t-$PtBi_2$. The expected peak positions for t-$PtBi_2$ are indicated by vertical bars (I) (ICSD No. 428088). The inset displays an image of the single crystal against a millimeter-sized grid. Right panel: SEM images recorded in secondary-electron (SE) and backscatter electron (BSE) modes, along with the EDX spectrum and chemical composition expressed in atomic percentage

**2.** Variety of *dV/dI* spectra of heterocontacts based on $PtBi_2$.

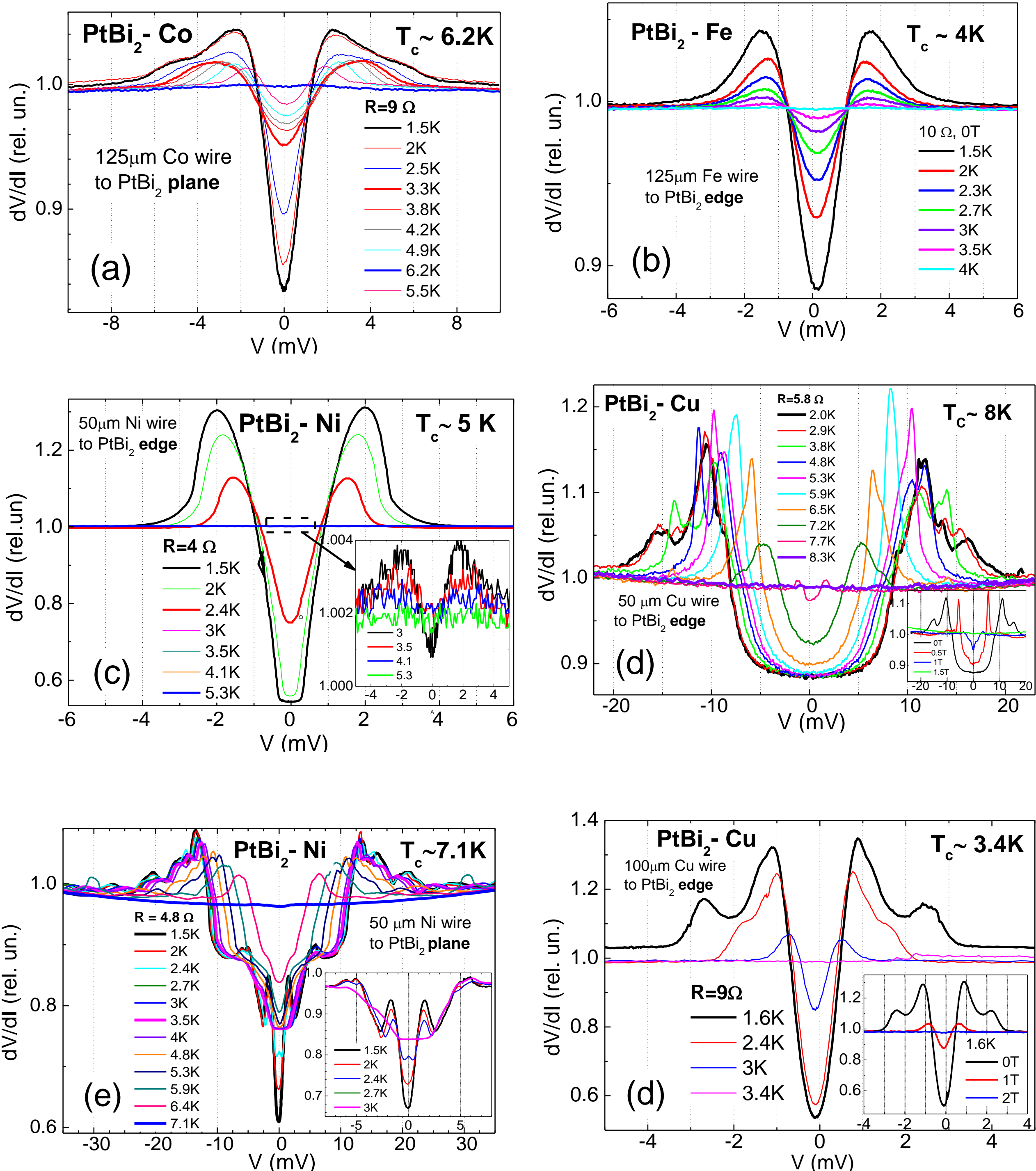

**Fig. S2**. Variety of *dV/dI* spectra of heterocontacts based on $PtBi_2$ with normal (Cu) and ferromagnetic metals (Fe, Co, Ni).

Figure S1 demonstrates the variety of *dV/dI* spectra of heterocontacts based on $PtBi_2$, both in shape and energy spread. Without delving into specifics or speculations, it is evident that the spectra do not display any consistent features or shapes that would be characteristic of Andreev reflection, as expected in superconductor-normal metal contacts under either ballistic or diffusive conditions [13]. Ubiquitous side peak(s), zero-bias dip, and large variation of the width of the whole structure, all this points to non-spectral regime in PCs. Therefore, the dV/dI features are most likely due to the critical current and/or heating effects in PC core.

**3**. Estimating the lower limit of the PC size.

Let us try to estimate the lower limit of the size of PCs from the resistance $R_{PC}$. The latter is expressed by the well-known Wexler formula [S1], which consists of the sum of ballistic Sharvin $R_{Sh}$ and diffusive Maxwell resistance $R_M$ :

$$R_{PC} = R_{Sh} + R_M \approx 16\rho l/3\pi d^2 + \rho/d, \qquad (1)$$

here d is the PC diameter, ρ is the resistivity, $l$ is the mean free path of electrons and $\rho l = p_F/ne^2 \approx 1.3\times10^4 n^{-2/3}[\Omega\cdot cm^2]$, where $p_F$ is the Fermi momentum, $n$ and e being the electron density per $cm^3$ and charge, respectively. Using the most common carrier density in $PtBi_2$ listed in the literature, $n\approx2.0\times10^{20}$ $cm^{-3}$, we obtained that $\rho l \approx 3.8\times10^{-10}$ $\Omega\cdot cm^2$. Here, in equation (1), we neglected the contribution of the counter-electrode, because $\rho l$ in elemental metals is at least order of magnitude less. Also, because ρ in PC is, in general, not well known due to surface contamination and possible disruption of the crystal structure in PC area by the tip, we will use only the Sharvin resistance to estimate of PC size. As a result, we obtain a lower limit for PC size between 45 and 150 nm for PCs in the range of 3 to 30 Ohm.